\documentclass[prl,twocolumn,preprintnumbers,amsmath,amssymb,superscriptaddress,floatfixX]{revtex4-1}

\usepackage{dcolumn}
\usepackage{graphicx}
\usepackage{epstopdf}
\usepackage{natbib}

\begin{document}
\title{A Single-Ion Nuclear Clock for Metrology at the 19$^{th}$ Decimal Place}

\author{C.\,J. Campbell}
\email{corey.campbell@gatech.edu}
\affiliation{
School of Physics, Georgia Institute of Technology, Atlanta, Georgia 30332-0430, USA}

\author{A.\,G. Radnaev}
\affiliation {
School of Physics, Georgia Institute of Technology, Atlanta, Georgia 30332-0430, USA}

\author{A. Kuzmich}
\affiliation {
School of Physics, Georgia Institute of Technology, Atlanta, Georgia 30332-0430, USA}

\author{ V.\,A. Dzuba}
\affiliation {
School of Physics, University of New South Wales, Sydney 2052, Australia}

\author{V.\,V. Flambaum}
\affiliation {
School of Physics, University of New South Wales, Sydney 2052, Australia}

\author{A. Derevianko}
\affiliation {
Department of Physics, University of Nevada, Reno, Nevada 89557, USA}

\date{October 7, 2011 }
\begin{abstract}
The 7.6(5)\,eV nuclear magnetic-dipole transition in a single $^{229}$Th$^{3+}$ ion may provide the foundation for an optical clock of superb accuracy.  A virtual clock transition composed of stretched states within the $5F_{5/2}$ electronic ground level of both nuclear ground and isomeric manifolds is proposed. It is shown to offer unprecedented systematic shift suppression, allowing for clock performance with a total fractional inaccuracy approaching 1$\times$$10^{-19}$.
\end{abstract}

\maketitle

The development of optical atomic clocks has produced tremendous advances in time and frequency metrology and their applications in tests of fundamental physics.  Various ionic and neutral atomic species are now utilized for frequency standards with fractional inaccuracies at or below $10^{-15}$ \cite{Tamm2009, Chwalla2009, Wilpers2007, Ludlow2008, Yamaguchi2011, Baillard2008, Akatsuka2010, Lemke2009, Poli2008}. The most accurate systems approach and even surpass the $10^{-17}$ mark \cite{Rosenband2008,Chou2010}.  However, the difficulty in simultaneous minimization of external field and motional shift uncertainties, as well as clock linewidth limitations, make the next two orders of magnitude of inaccuracy reduction challenging.  In 2003, Peik and Tamm pointed out that many of the standard external field shifts in atomic clocks could be suppressed or avoided in a $^{229}$Th$^{3+}$ optical nuclear transition by utilizing the $^2 S_{1/2}$ electronic level within both the nuclear ground and isomer manifolds \cite{Peik2003}.  This pioneering proposal illuminated a way towards a next-generation clock, however, their system is still subject to two important ion-clock limitations:  the proposed $m_F$\,=\,0\,$\leftrightarrow$\,$m_F$\,=\,0 clock transition experiences a significant second-order differential Zeeman shift ($\sim$\,70\,kHz/mT$^2$), while electric-quadrupole decay of both clock states limits the clock linewidth to $\sim$1\,Hz.

\begin{figure}[b]
\centering
\includegraphics[scale=.55,angle=90]{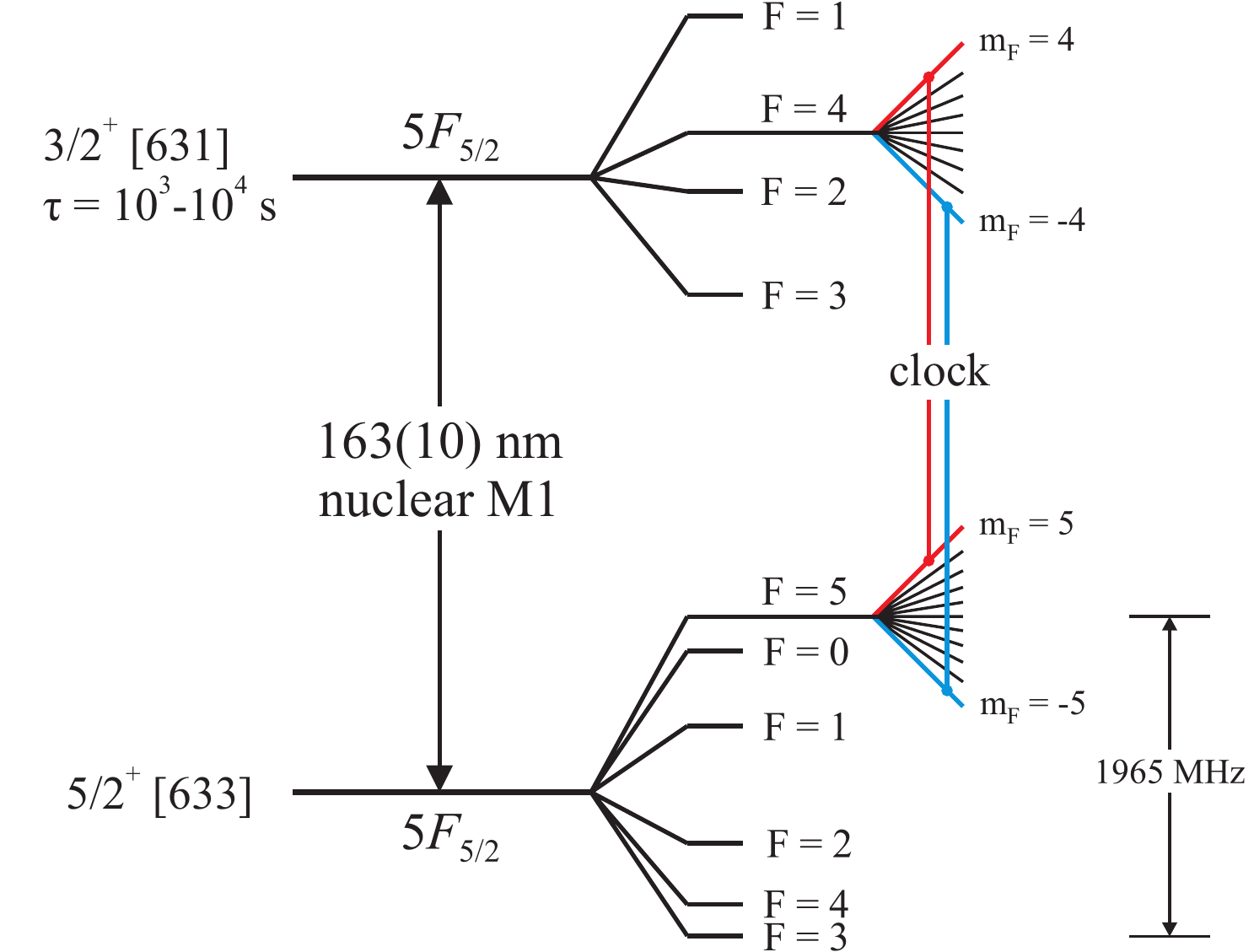}
\caption{Partial energy-level diagram of the $5F_{5/2}$ electronic ground levels within the nuclear ground and isomer manifold of $^{229}$Th$^{3+}$. The nuclear ground state hyperfine structure is taken from Ref.\cite{Campbell2011}, whereas the isomer hyperfine splittings are estimates.  See text for details.}
\label{fig:clock_transitions}
\end{figure}

In this Letter, we show that the pair of stretched hyperfine states within the $5F_{5/2}$ electronic ground level of both nuclear ground and isomeric manifolds in $^{229}$Th$^{3+}$ provide large suppression of all external field clock shifts, including the quadratic Zeeman shift, while providing a clock linewidth of $\lesssim$\,100\,$\mu$Hz.  The clock is composed of the $|5F_{5/2},$\,$I_g$\,=\,5/2;\,$F$\,=\,5,\,$m_F$\,=\,$\pm$5$\rangle$ $\leftrightarrow$ $|5F_{5/2}$,\,$I_m$\,=\,3/2;\,$F$\,=\,4,\,$m_F$\,=\,$\pm$4$\rangle$ transitions (Fig.~\ref{fig:clock_transitions}) which, when averaged within a weak magnetic field, form a virtual clock transition.  In general, stretched eigenstates in the coupled basis are also eigenstates in the uncoupled basis, i.e., $|F$\,=\,$I$+$J$,\,$m_F$\,=\,$\pm$$F\rangle$\,=\,$|J$,\,$\pm J\rangle$\,$\otimes$\,$ |I$,\,$\pm I\rangle$. For a structureless point-like nucleus, both ground and excited clock state shifts are then exactly equal.  When nuclear size and multipole moment differences in the clock levels are taken into account, differential shifts do arise, however they are suppressed well below levels of optical atomic clocks. This suppression encompasses shifts due to magnetic fields, electric fields, and electric field gradients, as described in detail below.  With such resilience to external-field perturbations, our proposal opens the door for metrology at a level unattainable by any other currently proposed clock.

Because the $5F_{5/2}$ orbitals are the electronic ground levels, clock interrogation time is fundamentally limited only by the $10^3$\,-\,$10^4$\,s nuclear isomer excited state lifetime~\cite{Dykhne1998, Ruchowska2006, Porsev2010}, providing potential for a resonator quality factor of $10^{19}$\,-\,$10^{20}$.  In a system where quantum projection noise dominates clock operation, the  Allan deviation is $\sigma_y(\tau)$\,$\approx$\,$(\gamma_{clk}/\nu_{clk})(\tau_{cycle}/\tau)^{1/2}$ \cite{Itano1993}, where $\gamma_{clk}$ is the clock excitation linewidth, $\nu_{clk}$ is the clock transition frequency, $\tau_{cycle}$ is the clock cycle time, and $\tau$ is the total integration time.  With an ultra-stable interrogation laser and efficient state detection allowing for $\gamma_{clk}$\,$\approx$\,1/$\tau_{cycle}$\,$\approx$\,1\,Hz, a $^{229}$Th$^{3+}$ clock at 1.8\,PHz could produce a quantum-limited fractional instability of 1$\times$$10^{-15} \,\tau^{-1/2}$, allowing for a statistical uncertainty of 1$\times$$10^{-18}$ to be achieved in 2\,weeks of averaging.  This conventional integration method may be used in conjunction with a VUV frequency comb~\cite{Jones2005, Ozawa2008, Yost2009} to compare the nuclear clock to an electronic anchor line in the same ion or to a $^{27}$Al$^+$ clock, for example, in order to probe for variation of fundamental constants at an unprecedented level of precision~\cite{Flambaum2006}.  With a laser-noise suppressing clock interrogation protocol \cite{Chou2011}, the long excited clock state lifetime could be more fully utilized, allowing for a pair of $^{229}$Th$^{3+}$ clocks to be compared with $\lesssim$\,$10^{-19}$ statistical uncertainty in only a few days of integration.

In conventional optical atomic clocks, the most significant external-field clock shifts arise due to differences in the clock state electronic wavefunctions.  In the nuclear system proposed here, all four clock state electronic wavefunctions are the same.  External-field shift analysis of this compound system (electron cloud $+$ nucleus) must then be applied to changes in nuclear properties and the resulting slight changes they create in the electronic wavefunction.  Consequently, the difference for external-field shift property $X$ (magnetic-dipole moment, electric-quadrupole moment, polarizability, etc...) comes in three flavors:
\begin{equation}
\Delta X{\mathcal{=\gamma}}_{X}\Delta X^{nuc}+\Delta X^{iso}+\Delta X^{hfs},
\label{eq:template}
\end{equation}
(i) $\Delta X^{nuc}$ is the direct nuclear contribution including core-electron shielding of externally-applied fields, ${\mathcal{\gamma}}_{X}$ being the shielding factor of Th$^{4+}$, (ii) $\Delta X^{iso}$ is the isomer nuclear-charge distribution (monopole-mediated) shift, and (iii) $\Delta X^{hfs}$ is the valence-electron contribution to the hyperfine-mediated shift.

To analyze contributions $\Delta X^{iso}$ and $\Delta X^{hfs}$, the electronic response of $^{229}$Th$^{3+}$ to changing nuclear properties is computed using techniques of relativistic many-body perturbation theory (MBPT).  The calculations start from the Dirac-Hartree-Fock approximation and incorporate the major correlation effect -- core polarization by the valence electron.  Calculations of matrix elements additionally include an all-order chain of random-phase-approximation diagrams. It is found that this approach recovers experimental hyperfine structure constants with an accuracy of 10\%, which, considering uncertainties in nuclear parameters, is sufficient for the goals of this Letter. Finally, various shielding factors for the Th$^{4+}$ core are taken from Ref.~\cite{FeiJoh69}.

To determine the monopole-mediated contribution, $\Delta X^{iso}$, the nuclear charge radius $R_{rms}$ is varied for a series of MBPT calculations of property $X$. This yields the rate of change $dX/dR_{rms}$. As a result, $\Delta X^{iso}$\,=\,$dX/dR_{rms} \Delta R_{rms}$, where $\Delta R_{rms}$ is the difference in the {\it rms} radii of the nuclear ground- and isomer-state charge distributions.  According to nuclear-structure calculations \cite{Litvinova2009}, $\left\vert \Delta R_{rms}\right\vert$\,$<$\,0.0038\,fm for the nuclear transition in $^{229}$Th.

The hyperfine-mediated shift is determined via first-order perturbation theory in the hyperfine interaction of electrons with the nuclear magnetic-dipole and electric-quadrupole moments. Generally, for the stretched states
\begin{equation}
\Delta X^{hfs}=\left(\frac{\mu_{m}-\mu_{g}}{\mu_N} \right) \bar{X}_{\mu}^{hfs}+\left(\frac{ \mathcal{Q}_{m}-\mathcal{Q}_{g}}{\left\vert e \right\vert b} \right) \bar{X}_{Q}^{hfs}\,,\label{Eq:Xhfs}
\end{equation}
where $\mu_{g,m}$ and $\mathcal{Q}_{g,m}$ are nuclear magnetic-dipole and electric-quadrupole moments of the ground ($g$) and isomer ($m$) states, and form-factors $\bar{X}_{\mu,Q}^{hfs}$ depend only on the electronic wavefunctions. Electronic form-factors are evaluated below using MBPT methods.  The nuclear M1 moments used are $\mu_{g}$\,=\,0.45\,$\mu_N$ \cite{Gerstenkorn1974} and $\mu_{m}$\,$\sim$\,-0.08\,$\mu_N$ \cite{Dykhne1998}.  The difference in intrinsic quadrupole moments of the two nuclear charge distributions, $\Delta\mathcal{Q}_{20}$, was investigated in Ref. \cite{Litvinova2009}. The condition $\left\vert \Delta \mathcal{Q}_{20} \right\vert$\,$<$\,0.28\,$\left\vert e\right\vert$fm$^2$ is taken from these results and, with $\mathcal{Q}_g$\,=\,3.11(16)\,$\left\vert e\right\vert$b \cite{Campbell2011}, the isomer quadrupole moment is $\mathcal{Q}_m$\,=\,1.74(17)\,$\left\vert e\right\vert$b.

Zeeman shifts arise due to interaction of external magnetic fields with the atomic magnetic dipole moment $\mu$.  In a weak magnetic field $\mathcal{B}$, the clock transition shifts are $\Delta E_Z^{(1)}$\,=\,$\pm \Delta \mu \mathcal{B}$, where in accordance with Eqn.\eqref{eq:template}, $\Delta \mu$\,=\,${\gamma}_{\mu}\Delta \mu^{nuc}$+$\Delta \mu^{iso}$+$\Delta \mu^{hfs}$.  The individual contributions are found to be ${\gamma}_{\mu}\Delta \mu^{nuc}$\,$\approx$\,-0.53\,$\mu_N$, $|\Delta \mu^{iso}|$\,$<$\,8$\times$$10^{-7}$\,$\mu_{N}$, and $\Delta \mu^{hfs}$\,$\approx$ \,-2$\times$$10^{-3}$\,$\mu_N$, which lead to differential Zeeman shifts of $\Delta E_Z^{(1)}/\mathcal{B}$\,$\approx$\,$\mp$4\,kHz/mT for the two clock transitions.

Due to the system's symmetry, only a magnetic field change which is correlated with alternation between clock transitions will produce a clock shift.  All other frequency components, e.g. ac line and $\mu$T-level fields at the trap drive frequency, will only modify the clock spectrum with small resolved modulation sidebands or slight line broadening.  The correlated field change should be held to $<$\,10\,pT at the ions to maintain a fractional shift of 1$\times$$10^{-20}$.  Magnetic shielding with $10^{-3}$ field suppression placed around the vacuum system could be combined with microwave spectroscopy, where the magnetic-field sensitivity enhancement is $\mu_B/\mu_N$\,$\sim$\,$1800$, to ensure this criterion is met.

The dominant contribution to second order Zeeman shifts is related to the magnetic-dipole polarizability of the nucleus itself. This quantity is enhanced due to the unusually small energy gap between the ground and isomeric states of the nucleus. As a result, the M1 polarizabilities of the two states have the same absolute value but opposite signs. The relevant clock shift may be estimated as $\Delta E_{Z}^{(2)}$ = $\left(2\,|\langle g~|M_{1}^{(n)}|~m\rangle |^2\, \mathcal{B}^2 \right)/\Delta E_{clk}$\,$\sim$\,$\left(2\mu_{N}^{2}\mathcal{B}^{2}\right)/\Delta E_{clk}$.  A 1\,$\mu$T field fluctuation at the clock ion, e.g. from rf currents in the trap electrodes, would then lead to a fractional clock shift of only $\sim$\,4$\times$$10^{-29}$.

The quadrupole shift arises due to interaction of the atomic quadrupole moment $\mathcal{Q}$ with external electric field gradients.  The value of $\gamma_{Q}\Delta\mathcal{Q}^{nuc}$\,$\approx$\,9$\times$$10^{-6}$\,$\left\vert e\right\vert a_{0}^{2}$ is estimated using $\Delta \mathcal{Q}^{nuc}$\,=\,-1.37(5)\,$\left\vert e\right\vert$b discussed above and the quadrupole shielding factor $\gamma_{Q}$\,=\,-177.5 for Th$^{4+}$ \cite{FeiJoh69}.  The monopole-mediated shift is calculated to be $|\Delta\mathcal{Q}^{iso}|$\,$<$\,8$\times$$10^{-7}$\,$\left\vert e\right\vert a_{0}^{2}$. Electronic form-factors entering the hyperfine-mediated shift, Eqn.\eqref{Eq:Xhfs}, are found to be $\bar{Q}_{\mu}^{hfs}$\,=\,-2.0$\times$$10^{-6}$\,$\left\vert e\right\vert a_{0}^{2}$ and $\bar{Q}_{Q}^{hfs}$\,=\,-5.1$\times$$10^{-7}$\,$\left\vert e\right\vert a_{0}^{2}$, leading to $\Delta \mathcal{Q}^{hfs}$\,$\approx$\,1.8$\times$$10^{-6}$\,$\left\vert e\right\vert a_{0}^{2}$.  Combining all three sources of the quadrupole clock shift, $\Delta\mathcal{\mathcal{Q}}$\,$\approx$\,1$\times$$10^{-5}$\,$\left\vert e\right\vert a_{0}^{2}$, a result dominated by the anti-screened direct nuclear contribution.

The dominant electric field gradient at the clock ion is due to the dc electric field which produces axial ion confinement within a linear rf trap.  Using an axial secular frequency of 500 kHz, the resultant 8$\times$$10^6$\,V/m$^2$ gradient would produce a fractional frequency shift of $\approx$\,3$\times$$10^{-20}$, a value sufficiently small to avoid clock averaging over three orthogonal quantization axes \cite{Itano2000}.

The Stark shift arises via atomic polarization induced by external electric fields.  With an exceedingly small differential nuclear polarizability ($\Delta \alpha^{nuc}$\,$<$\,$10^{-14}$\,$a_0^3$) and efficient screening of dc electric fields at the nucleus by the electron cloud, the direct nuclear contribution to static Stark shifts is found to be negligible. Using MBPT, the static scalar ($s$) and tensor ($t$) electronic polarizabilities are evaluated for the $5F_{5/2}$ state, resulting in $\alpha_{s}\approx 13\,a_{0}^{3}$ and $\alpha_{t}$\,$\approx$\,-4.2\,$a_{0}^{3}$.  From these results with varied nuclear size, $|\Delta\alpha^{iso}|$\,$<$\,3$\times$$10^{-5}$\,$a_{0}^{3}$.

Calculation of the hyperfine-mediated shift, Eqn.\eqref{Eq:Xhfs}, is done using the formalism and codes in Ref. \cite{Dzuba2010}.  The resulting form-factors are, in units of $a_{0}^{3}$: $[\bar{\alpha}_{\mu}^{hfs}]_{s}$\,=\,-4.9$\times$$10^{-6}$, $[\bar{\alpha}_{\mu}^{hfs}]_{t}$\,=\,1.4$\times$$10^{-5}$, $[\bar{\alpha}_{Q}^{hfs}]_{s}$\,=\,7.1$\times$$10^{-6}$, and $[\bar{\alpha}_{Q}^{hfs}]_{t}$\,=\,-9.3$\times$$10^{-6}$. With these values, the HFI-mediated differential polarizability is $\Delta\alpha^{hfs}$\,$\approx$\,-1$\times 10^{-5}\,a_{0}^{3}$ and, finally, $\left\vert \Delta\alpha\right\vert$\,$<$\,4$\times$$10^{-5}$\,$a_{0}^{3}$, dominated by the electric monopole-mediated contribution.

The fractional blackbody radiation (BBR) clock shift is parameterized as $\delta\nu_{BBR}/\nu_{clk}$\,=\,$\beta$$\times$$\left( T/T_0 \right)^{4}$, where $T_0$\,=\,300\,K.  In atomic units, $\beta$\,=\,-$\left[(\alpha^{3}\pi^{2} T_0 ^{4})/(15\,\nu_{clk})\right]$$\times$$\left[  \Delta\alpha\right]_{s}$, where $\left[  \Delta\alpha\right]  _{s}$ is the difference in the static scalar polarizabilities of the two clock states. From the static polarizability presented above, $\left\vert \beta^{iso}\right\vert$\,$<$\,1$\times$$10^{-22}$ and $\beta^{hfs}$\,$\approx$\,3$\times$$10^{-23}$, with the direct nuclear contribution being negligibly small.  The BBR shift is dominated by the electric monopole-mediated contribution and translates into a fractional clock shift of $<$\,1.3$\times$$10^{-22}$ at room temperature, several orders of magnitude below BBR shifts in current generation atomic clocks.

Interrogation with the clock laser will itself induce a differential light shift on the clock states.  Because the nuclear differential polarizability is exceedingly small, a direct nuclear contribution is negligible. For a circularly-polarized interrogation field, $\Delta\alpha^{iso}(\omega_{clk})$\,$\approx$\,-5.3$\times$$10^{-6}$\,$a_0^3$ within the expected clock transition energy range of 6.1-9.1\,eV.  Similarly, the hyperfine-mediated contribution remains nearly constant at $\Delta\alpha^{hfs}(\omega_{clk})$\,$\approx$\,-1.6$\times$$10^{-7}$\,$a_0^3$, leading to $\Delta\alpha(\omega_{clk})$\,$\approx$\,-5.3$\times$$10^{-6}$\,$a_0^3$.

To evaluate the interrogation laser light shift, a lower estimate of 0.2\,$\mu_N$ is assumed for the clock transition matrix element.  A cw laser intensity of 10\,mW/cm$^2$ would then induce a clock Rabi frequency of $\sim$\,1\,Hz and result in a fractional light shift of $\sim$\,1$\times$$10^{-24}$.

A room-temperature vacuum system would house the clock ion trap and could be maintained at a pressure of 1$\times$$10^{-10}$\,Pa.  With this low background pressure and extremely small differential polarizability of the clock, background-gas collisional shifts are expected to be negligible, though our estimate allows for an uncertainty of 1$\times$$10^{-20}$.

In addition to a narrow clock transition and sizable external-field shift suppression, the $^{229}$Th$^{3+}$ clock contains properties favorable for significant reduction of second-order motional (time dilation) shifts, a dominant systematic in ion clocks.  The 150\,kHz wide $5F_{5/2}$\,$\leftrightarrow$\,$6D_{3/2}$ electric dipole transition at 1088\,nm allows for direct ground-state cooling to minimize motion while thorium's large mass further reduces the time dilation shift.  To maintain motional ground state occupation throughout clock interrogation/evolution, a $^{232}$Th$^{3+}$ ion may be used for sympathetic cooling.  The cooling field would overlap both ions, cooling the 232 isotope, while an additional 1088\,nm compensation field would serve to mitigate differential light shifts of the clock states.

A pair of Th$^{3+}$ ions would be stored in a linear rf trap where strong rf confinement places both ions on the trap axis ($z$-axis).  Within a trap of radius 700\,$\mu$m and drive frequency of 25\,MHz, the four transverse motional eigenfrequencies may be set to 2.0\,-\,2.5\,MHz with the two axial frequencies at 0.5\,-\,0.9\,MHz.  In this case, sideband cooling well into the ground state could be achieved for all modes directly with a 1088\,nm field \cite{Peik1999,Rohde2001}. The total kinetic energy of the clock ion is then $E_K$\,$\approx$\,$\tfrac{1}{4} h ( \nu_{x0}$+$\nu_{x\pi}$+$\nu_{y0}$+$\nu_{y\pi}$)\,+\,$\tfrac{1}{8} h ( \nu_{z0}$+$\nu_{z\pi}$), where $\nu_{i\theta}$ is the secular frequency along dimension $i$ with relative ion motional phase $\theta$.  Time dilation results in a fractional clock frequency shift of $\delta \nu_{TD}/\nu_{clk}$\,=\,-$E_K/mc^2$, where $m$ is the ion mass. For $E_K$ associated with ground-state secular motion and minimal micromotion, $\delta \nu_{TD}/\nu_{clk}$\,$\approx$\,-5$\times$$10^{-20}$.  With motional state measurements of $\langle n_{i \theta} \rangle$\,$\ll$\,$\tfrac{1}{2}$ for all modes, the uncertainty of this shift could be held to $<$\,1$\times$$10^{-20}$.

A second contribution to time dilation comes from excess micromotion (EMM) \cite{Berkeland1998} arising from stray electric fields which push the clock ion away from the rf potential minimum.  In the pseudopotential approximation, EMM kinetic energy due to a stray electric field $\mathcal{E}_i$ along radial dimension $i$ may be estimated as $E_{EMMi}$\,$\approx$\,(2$m$)$^{-1}$$( q \, \mathcal{E}_i/2 \pi \nu_{i0})^2$, where $q$ is the ion charge.  If stray electric fields along the $x$- and $y$-dimensions are confined to $\lesssim$\,1\,V/m with proper trap design and continuous motional monitoring, the additional contribution to time dilation from excess micromotion would be $\lesssim$\,$1.0\times$$10^{-19}$.  The uncertainty of this systematic shift is taken as the shift upper estimate itself.

The electric fields used for ion confinement will induce a small differential Stark shift on the clock states \cite{Berkeland1998}.  For the ground state cooled system with a 1\,V/m transverse electric field, a shift is estimated from the results above to be $<$\,3$\times$$10^{-26}$.

A single $\sigma^{\pm}$ transition at 1088\,nm is dipole allowed from each of the four clock states.  A light shift of the $i^{th}$ clock state due to coupling of this transition to the $^{232}$Th$^{3+}$ cooling field is $\Delta \nu_{i1088}$\,$\approx$\,$\Omega_{R}^2/(4 \Delta_i)$, where $\Omega_{R}$ is the resonant Rabi frequency induced between stretched states within the cooling ion and $\Delta_i$ is the detuning of the cooling field from the relevant 1088\,nm transition in the clock ion.  This detuning is dominated by the 10\,GHz isotope shift \cite{Campbell2011}.  Though the dipole matrix elements of all four 1088\,nm transitions in the clock ion are equal, the leading differential light shift arises via different values of $\Delta_i$, due primarily to the $\lesssim$\,1.4\,GHz isomer field shift \cite{Berengut2009} and secondarily to the $\sim$\,100\,MHz differential hyperfine shift.  The differential shift of the clock transition is then $\lesssim$\,15\% of the nuclear ground clock state light shift.  Using $\Omega_{R}$\,=\,150\,kHz for ground state cooling equates to a clock shift of $\lesssim$\,85\,mHz.

An additional 1088\,nm field can be used for Stark compensation, with the leading limitation in suppression due to uncorrelated intensity fluctuations between the cooling and compensation field.  With $10^{-3}$ fractional intensity instability between the two fields, differential Stark shifts can be suppressed by three orders of magnitude, leading to a clock shift and uncertainty of 5$\times$$10^{-20}$.  Off-resonant scattering from the clock states due to the 1088\,nm cooling and compensation fields is estimated to produce $<$\,10\,mHz of clock transition broadening.

To remove linear Doppler shifts which are correlated with clock interrogation, alternating clock probes from counter-propagating directions may be averaged \cite{Rosenband2008}.  By minimizing changes in optical intensities, polarizations, and detunings when alternating between probe directions, this differential linear Doppler shift can be held to 1 $\times$$10^{-17}$ \cite{Chou2010}.  With an asymmetry of the clock lineshapes and respective servo gains held to 0.1$\%$, the fractional linear Doppler uncertainty can then be confined to 1$\times$$10^{-20}$.

Gravitational shifts will play a significant role in clocks operating with inaccuracy near 1$\times$$10^{-19}$.  The fractional frequency difference between two clocks at differing heights on Earth's surface is $\Delta \nu_G/\nu_{clk}$\,=\,$g \, \Delta h/c^2$, where $g$ is the mean gravitational acceleration at the clocks and $\Delta h$ is the clock height difference.  This results in a 1$\times$$10^{-19}$ fractional frequency shift for a 1\,mm discrepancy in clock height, a likely level of error for two separated clocks in the same lab.

\begin{table}\small
\caption{\label{tab:table1} Estimated systematic error budget for a $^{229}$Th$^{3+}$ clock using realized single-ion clock technologies.  Shifts and uncertainties are in fractional frequency units ($\Delta \nu / \nu_{clk}$) where $\nu_{clk}$ = 1.8\,PHz.  See text for discussion.}
\begin{ruledtabular}
\begin{tabular}{lcc}
\textrm{Effect}&
\textrm{$\vert$Shift$\vert$ (10$^{-20}$)}&
\textrm{Uncertainty (10$^{-20}$)}\\
\colrule
Excess micromotion & 10 & 10 \\
Gravitational & 0 & 10\\
Cooling laser Stark & 0 & 5 \\
Electric quadrupole & 3 & 3 \\
Secular motion & 5 & 1 \\
Linear Doppler & 0 & 1 \\
Linear Zeeman & 0 & 1 \\
Background collisions & 0 & 1 \\
Blackbody radiation & 0.013 & 0.013 \\
Clock laser Stark & 0 & $\ll$\,0.01 \\
Trapping field Stark & 0 & $\ll$\,0.01 \\
Quadratic Zeeman & 0 & 0 \\
Total & 18 & 15
\end{tabular}
\end{ruledtabular}
\end{table}
A large set of clock shifts relevant to the proposed system has been analyzed, with an error budget including these shifts and uncertainties found in Table \ref{tab:table1}.  With all classes of shift uncertainty added in quadrature, the total estimated clock inaccuracy is 1.5$\times$$10^{-19}$.

Utilization of the $^{229}$Th nuclear transition for high-accuracy frequency metrology would require ultra-stable interrogation light at 163(10)\,nm.  With a $\mu$W-level cw power requirement, such light might be generated by multi-stage sum-frequency generation of visible or NIR light within nonlinear crystals, e.g. LBO, BBO, and KBBF, with the fundamental frequency locked to an ultra-stable optical resonator.  To determine the state of the clock after interrogation, a double-resonance method may be used where NIR or visible fluorescence from an electric-dipole transition in the clock ion is correlated with a particular nuclear state \cite{Peik2003}.  Due to the long isomer state lifetime, the clock interrogation protocol could begin from both the nuclear ground (excitation) and isomer (de-excitation) manifolds to expedite the clock initialization process.

In conclusion, this work has shown that the usual sources of clock shifts are suppressed in the proposed system, so that clock operation with inaccuracy approaching  1$\times$$10^{-19}$ appears viable.  With a more aggressive pursuit of systematic shift compensation and suppression, this value may very well be lowered to the $10^{-20}$ scale, however,  other systematic clock shift mechanisms may become operational at this level. 

This work was supported by the Office of Naval Research, the National Science Foundation, and the Gordon Godfrey fellowship.

\end{document}